\documentclass[preprint]{aastex}
\slugcomment{Accepted for publication in The Astrophysical Journal}

\shorttitle{Warping and precession in accretion disks}
\shortauthors{Caproni et al.}

\begin{document}


\title{Warping and precession in galactic and extragalactic accretion disks}


\author{Anderson Caproni\altaffilmark{1,2}, Mario Livio\altaffilmark{2}, Zulema Abraham\altaffilmark{1} and Herman J. Mosquera Cuesta \altaffilmark{3,4}}

%%\and

%%\author{Herman J. Mosquera Cuesta \altaffilmark{1,2}}

\altaffiltext{1}{Instituto de Astronomia, Geof\'\i sica e Ci\^encias
Atmosf\'ericas, Universidade de S\~ao Paulo, R. do
Mat\~ao 1226, Cidade Universit\'aria, CEP 05508-900, S\~ao Paulo, SP,
Brazil; acaproni@astro.iag.usp.br.}

\altaffiltext{2}{Space Telescope Science Institute, 3700 San Martin Drive, Baltimore, MD, 21218.}

\altaffiltext{3}{Instituto de Cosmologia, Relatividade e Astrof\'{\i}sica 
(ICRA-BR),  Centro Brasileiro de Pesquisas F\'\i sicas, R. Dr. Xavier 
Sigaud 150, CEP 22290-180, Rio de Janeiro, RJ, Brazil; hermanjc@cbpf.br.}

\altaffiltext{4}{Abdus Salam International Centre for Theoretical Physics, 
Strada Costiera 11, Miramare 34014, Trieste, Italy}

%%\email{aastex-help@aas.org}

%%\and

%%\author{R. J. Hanisch\altaffilmark{5}}
%%\affil{Space Telescope Science Institute, Baltimore, MD 21218}


%%\altaffiltext{2}{Society of Fellows, Harvard University.}
%%\altaffiltext{3}{present address: Center for Astrophysics,
%%    60 Garden Street, Cambridge, MA 02138}
%%\altaffiltext{4}{Visiting Programmer, Space Telescope Science Institute}
%%\altaffiltext{5}{Patron, Alonso's Bar and Grill}


\begin{abstract}
   The Bardeen-Petterson general relativistic effect has been suggested 
as the mechanism responsible for precession in some accretion disk systems. 
Here we examine separately four mechanisms (tidally-induced, irradiation-induced, 
magnetically-induced and Bardeen-Petterson-induced) that can lead to 
warping and precession. We use a sample of eight X-ray binaries 
and four Active Galactic Nuclei (AGNs) that present signatures of 
warping and/or precession in their accretion disks to explore the viability 
of the different mechanisms. For the X-ray binaries SMC\,X-1 and 4U\,1907+09 
all four mechanisms provide precession periods compatible with those observed, 
while for Cyg\,X-1 and the active galaxies Arp\,102B and NGC\,1068, 
only two mechanisms are in agreement with the observations. 
The irradiation-driven instability seems incapable of producing 
the inferred precession of the active galaxies in our sample, and the 
tidally-induced precession can probably be ruled out in the case 
of Arp\,102B. Perhaps the best case for a Bardeen-Petterson precession 
can be achieved for NGC\,1068. Our results show that given the many 
observational uncertainties that still exist, it is extremely difficult 
to confirm unambiguously that the Bardeen-Petterson effect has been 
observed in any of the other sources of our sample.
\end{abstract}


%% subject header.  Objects should be in the appropriate "individual"
%% headers (e.g. quasars: individual, stars: individual, etc.) with the
%% additional provision that the total number of headers, including each
%% individual object, not exceed six.  The \objectname{} macro, and its
%% alias \object{}, is used to mark each object.  The macro takes the object
%% name as its primary argument.  This name will appear in the paper
%% and serve as the link's anchor in the electronic edition if the name
%% is recognized by the data centers.  The macro also takes an optional
%% argument in parentheses in cases where the data center identification
%% differs from what is to be printed in the paper.



\section{Introduction}

   The importance of accretion processes in the line formation and 
continuum emission in AGNs and galactic sources, such as microquasars and 
X-ray binaries, has long been recognized in the literature, even though 
the presence of an accretion disk in some of those systems is inferred 
only indirectly.

   The continuous improvement in the capability of telescopes and 
detectors to obtain high-resolution images, and spectra with higher 
sensitivity, allowed for the direct probing of the physical characteristics 
of some accretion disk systems (e.g., \citealt{ray96,jon00,jaf04,fat06}).

   In contrast to the standard picture of a flat disk surrounding 
the accreting object, some galactic and extragalactic sources present 
signatures of warping and precession in their disks and/or jets (e.g., 
\citealt{marg84,sil88,abra00,gal04,cap06}). The apparent lack of 
correlation between the orientation of the radio jets and the plane 
of the host galaxy's disk might also be attributed to warped disks 
(e.g., \citealt{sch02}). 

   The Bardeen-Petterson effect \citep{bape75}, a gravitational 
perturbation predicted by general relativity, is one of the physical 
mechanisms that has been proposed to explain warping and precession 
in accretion disks (e.g., \citealt{scfe96,nepa00,fran05,kin05,cap06}). 
In order to explore whether this mechanism is universal, in the sense that 
it can be acting in both galactic and extragalactic accretion disks, 
we have selected eight X-ray binaries and four AGNs that present 
signatures of warping/precession in their disks. We have analyzed 
individually the predictions of three additional precession mechanisms 
for the same sample of sources, in order to compare these predictions 
with those from the Bardeen-Petterson effect. In Section 2, we 
describe the accretion disk model used throughout the paper, as well 
as constraints on the basic parameters of the accretion disk. In 
Section 3, we present the four warping/precession mechanisms studied 
in this work. A brief introduction of the sample of sources, as 
well as their basic parameters (inferred observationally) is given 
in Section 4. A comparison between the theoretical predictions and 
observations is carried out in Section 5, and conclusions follow.

\section{The accretion disk model}

   We consider an accretion disk through which mass is accreted 
onto a compact object (a black hole or a neutron star) at a mass 
accretion rate $\dot{M}$. The angular momentum of the accretion 
disk per unit area can be written as:

\begin{eqnarray}
   L_\mathrm{d}(r) = \Sigma(r)\Omega(r)r^2,
\end{eqnarray}
\\where $r$ is the radial distance from the compact object, $\Sigma$ 
is the surface density of the accretion disk (integrated over a scale 
height $H$), and $\Omega$ is the (relativistic) Keplerian angular velocity 
of the disk, given by (e.g., \citealt{abr78}):

\begin{eqnarray}
   \Omega_\mathrm{K}(r) = \frac{c}{R_\mathrm{g}}\left[\left(\frac{r}{R_\mathrm{g}}\right)^{3/2}+a_\ast \right]^{-1},
\end{eqnarray}
\\where $a_\ast$ is the ratio between the actual angular momentum of 
the compact object and its maximum possible value and $R_\mathrm{g}=GM/c^2$ 
is the gravitational radius of the compact object, (where $G$ and $c$ 
are the gravitational constant and the speed of light respectively). Note that for 
$r\gg R_\mathrm{g}$ or $a_\ast\simeq 0$, we recover the Newtonian angular 
velocity.

   If the accretion disk is not self-gravitating, the scale height of 
the disk can be determined from hydrostatic equilibrium to be:

\begin{eqnarray}
  H(r) = \frac{c_\mathrm{s}(r)}{\Omega(r)},
\end{eqnarray}
\\where $c_\mathrm{s}$ is the sound speed, given by (e.g., \citealt{abr88}):

\begin{eqnarray}
   c_\mathrm{s}(r) = \sqrt{-\frac{5}{3}\frac{d\ln\Omega(r)}{d\ln r}
    \frac{\nu_1(r)\Omega(r)}{\alpha}},
\end{eqnarray}
\\where $\alpha$ is the dimensionless viscosity parameter introduced by 
\citet{shsu73}. The azimuthal kinematic viscosity of the disk, 
$\nu_1$, is calculated from (e.g., \citealt{krol98}):

\begin{eqnarray}
   \nu_1(r) = -\frac{\dot{M}}{2\pi\Sigma(r)}\left[
     \frac{d\ln\Omega(r)}{d\ln r}\right]^{-1}\left[1-\left(\frac{R_\mathrm{ms}}{r}\right)^2
     \frac{\Omega(R_\mathrm{ms})}{\Omega(r)}\right],
\end{eqnarray}
\\where $R_\mathrm{ms}$ is the radius of the innermost marginally stable orbit, 
assumed to be the inner radius of the disk\footnote{For neutron stars 
with strong magnetic fields, the accretion disk might be truncated at 
a larger radius than $R_\mathrm{ms}$, at the magnetospheric radius. 
Even though this might be the case for some sources in our sample, it 
should not change our results in any substantial way.}.

   In this work, we assume a power-law disk (e.g., 
\citealt{ost92,mal98,cap04,ray05,cap06}):

\begin{eqnarray}
   \Sigma(r) = \Sigma_0\left(\frac{r}{R_\mathrm{g}}\right)^s,
\end{eqnarray}
\\where $\Sigma_0$ and $s$ are constants.

   For accretion disk models available in the literature, 
$-2<s<2$, while $\Sigma_0$ should be determined from some reasonable 
assumptions concerning each system.

   An extreme upper limit for $\Sigma_0$ can be found by imposing that 
the accretion disk mass $M_\mathrm{d}$ is lower than the mass of the 
compact object, such that

\begin{eqnarray}
   \Sigma_0^{M_\mathrm{d}} < \frac{c^2}{2\pi G} R_\mathrm{g}^{s+1}\left(\int^{R_\mathrm{out}}_{R_\mathrm{ms}}r^{s+1}dr\right)^{-1}.
\end{eqnarray}

   In order to estimate the limit on $\Sigma_0$ using equation (7), it is necessary 
to know the outer radius of the disk $R_\mathrm{out}$, which is generally 
not well constrained by observations, especially for AGNs. For our AGN sample, 
we have adopted $R_\mathrm{out}\sim 10^5 R_\mathrm{g}$ based on the model of 
\citet{codu90}, except for NGC\,1068 in which interferometric maser observations 
indicate that its outer radius is at about 1.1 pc \citep{gal04}. In the case of 
binary systems in which accretion occurs via Roche lobe overflow, we take 
$R_\mathrm{out}\approx 0.88 R_\mathrm{L}$ \citep{papr77}, where the Roche-lobe 
radius $R_\mathrm{L}$ is given by \citep{eggl83}:

\begin{eqnarray}
   R_\mathrm{L} = \frac{0.49q^{2/3}}{0.6q^{2/3}+\ln(1+q^{1/3})}R_\mathrm{ps},
\end{eqnarray}
\\where $R_\mathrm{ps}$ is the binary separation.

   Another estimate for $\Sigma_0$ can be derived from the assumption that 
the accretion disk is self-gravitationally stable. This implies that the Toomre 
parameter $Q = c_\mathrm{s}\Omega/(\pi G\Sigma)$ \citep{toom64} must be greater 
than unity, which leads to:

\begin{eqnarray}
   \Sigma_0^{Q} < \left[\frac{5c^3}{6\pi^3G^2R_\mathrm{g}^3}\frac{\dot{M}}{\alpha}\right]^{1/3}\left(\frac{R_\mathrm{out}}{R_\mathrm{g}}\right)^{-(s+3/2)}F_\mathrm{Q}^{1/3},
\end{eqnarray}
\\with the dimensionless function $F_\mathrm{Q}$ given by:

\begin{eqnarray}
   F_\mathrm{Q}(a_\ast,R_\mathrm{out}) = \left[\frac{1-\left(R_\mathrm{ms}/R_\mathrm{out}\right)^2\Omega(R_\mathrm{ms})/\Omega(R_\mathrm{out})}{1+a_\ast\left(R_\mathrm{out}/R_\mathrm{g}\right)^{-3/2}}\right].
\end{eqnarray}

   We can see that $F_\mathrm{Q}\sim 1$ for $R_\mathrm{out}\gg R_\mathrm{ms}$, 
which is true for all the sources in our sample. The upper limit on $\Sigma_0$ 
corresponds to the minimum between $\Sigma_0^{M_\mathrm{d}}$ and $\Sigma_0^{Q}$.

   On the other hand, a lower limit for $\Sigma_0$ can be established from the 
requirement that the radial inflow is (highly) subsonic. This gives:

\begin{eqnarray}
   \Sigma_0^{v_\mathrm{r}} > \frac{1}{\Upsilon^2}\frac{3\alpha\dot{M}}{10\pi cR_\mathrm{g}}\left(\frac{R_\mathrm{out}}{R_\mathrm{g}}\right)^{-(s+1/2)}\left[1+a_\ast\left(\frac{R_\mathrm{out}}{R_\mathrm{g}}\right)^{-3/2}\right]^{-1}F_\mathrm{Q}.
\end{eqnarray}
\\where the Mach number $\Upsilon=v_\mathrm{r}/c_\mathrm{s}$, and $v_\mathrm{r}$ 
is radial velocity of the disk material. Although the requirement of a subsonic 
accretion inflow only implies that $\Upsilon<1$, current accretion disk models 
usually give $\Upsilon\lesssim 0.01$ (e.g., \citealt{shsu73,abr88,nayi95}). 
Consequently, we have assumed $\Upsilon=0.01$ in this work.

   We therefore take for the allowed range of $\Sigma_0$ in this work: 
$\Sigma_0^{v_\mathrm{r}} < \Sigma_0 < \mathrm{min}\left(\Sigma_0^{M_\mathrm{d}},\Sigma_0^\mathrm{Q}\right)$. 
We should note that for the irradiation-driven instability, another lower 
limit on $\Sigma_0$ can be derived from considerations of the disk opacity 
(see $\S$ 3.2). However, in most of the cases analyzed in this work, 
$\Sigma_0^{v_\mathrm{r}}$ has provided a more restrictive lower limit for 
$\Sigma_0$.

\section{Warp/Precession mechanisms}

   There are four main mechanisms that have been suggested for driving warping 
and precession in accretion disks. We consider all of these in turn.

\subsection{Tidal forces of a companion object in a binary system}

   The precession of an accretion disk can be tidally induced by a 
companion in a binary system (e.g., \citealt{katz73,sil88,katz97,rom00,caab04a}). 

   We consider a binary system with masses $M_\mathrm{p}$ and $M_\mathrm{s}$ 
(for the primary and secondary, respectively), separated by a distance 
$R_\mathrm{ps}$. From Kepler's third law,

\begin{eqnarray}
   R_\mathrm{ps}^3 = \frac{\mathrm{G}(M_{\mathrm{p}}+M_{\mathrm{s}})}{4\pi^2}P_\mathrm{orb}^2,
\end{eqnarray}
\\where $P_\mathrm{orb}$ is the orbital period.

   If the orbit is non-coplanar with the accretion
disk, torques are induced in the inner parts of the disk, producing 
precession. Taking the outer radius of the precessing disk to be 
$R_\mathrm{prec}$ ($R_\mathrm{prec}\leq R_\mathrm{out}$; \citealt{rom00}), 
the precession period, $P_\mathrm{prec}$, is given by \citep{pate95,larw97}:

\begin{eqnarray}
  P_{\mathrm{prec}} \geq -\frac{8\pi}{3}\left(\frac{5-n}{7-2n}\right)\frac{R_{\mathrm{ps}}^3}{R_{\mathrm{prec}}^{3/2}}\frac{1}{\sqrt{GM_{\mathrm{p}}}q\cos\theta},
\end{eqnarray}
\\where $n$ is the polytropic index of the gas (e.g., $n=3/2$ for a non-relativistic 
gas and $n=3$ for the relativistic case), $q=M_{\mathrm{s}}/M_{\mathrm{p}}$ and $\theta$ 
is the inclination angle of the orbit with respect to the disk plane. The negative 
sign in equation (13) indicates that the induced precession is retrograde, in the 
sense of being contrary to the rotation of the accretion disk.

   Combining equations (12) and (13) (with the condition $R_\mathrm{prec}\leq R_\mathrm{out}$) 
we obtain:

\begin{eqnarray}
  P_\mathrm{prec} \geq -\frac{4}{3}\left(\frac{5-n}{7-2n}\right)\left[\frac{(1+q)^{1/3}}{0.88q^{2/3}f(q)}\right]^{3/2}\frac{P_\mathrm{orb}}{\cos\theta},
\end{eqnarray}
\\where $f(q)$ is the function multiplying $R_\mathrm{ps}$ on the 
right-hand side of equation (8). Note that the ratio between the 
precession and orbital periods depends (apart from $\theta$) 
primarily on the mass ratio of the binary system.

\subsection{Radiation-driven instability}

   Radiation produced by the accreting compact object can modify 
the dynamics of an optically thick accretion disk. \citet{pett77} 
showed that if the disk is warped and optically thick, the radiation 
pressure will produce nonaxisymmetric torques that will change the 
initial warped configuration of the disk. \citet{prin96} and subsequently 
\citet{mal96} showed that even for an initially flat accretion 
disk, the radiation torques can warp and twist it. \citet{prin97} 
extended those calculations to include self-shadowing effects 
due to the warps. This radiation-driven instability has been invoked 
to explain anomalies in the morphology and variability in a huge 
variety of astrophysical sources (e.g., 
\citealt{cli95,mal96,sou97,lipr97,arpr97,wipr99,ogdu01}).

   \citet{mal98} generalized the isothermal disk model assumed in 
previous papers, and considered power-law surface density distributions 
ranging from the isothermal case ($s=-3/2$) to radiation-pressure 
dominated disks ($s=3/2$). Solving numerically the twist differential 
equation governing such disks (without including self-shadowing), 
they found that the critical radius at which the disk becomes 
unstable is:

\begin{eqnarray}
   R_\mathrm{cr} = x_\mathrm{cr}^2\left(\frac{\eta}{\epsilon}\right)^2R_\mathrm{g},
\end{eqnarray}
\\where $\eta=\nu_2/\nu_1$ is the ratio between the vertical and the 
azimuthal viscosities in the disk, $\epsilon$ is the accretion efficiency 
and $x_\mathrm{cr}$ is equal to $2\pi$ for $s=-3/2$, increasing 
monotonically to $\sim 4.891\pi$ for $s=3/2$.
   
   In the steady-state regime, the precession period, $P_\mathrm{prec}^\mathrm{rad}$, 
due to the radiation-instability is \citep{mal98}: 

\begin{eqnarray}
   P_\mathrm{prec}^\mathrm{rad} = \frac{4\pi^2}{\sigma_0}\tilde{\sigma}^{-1}\left(\frac{\eta}{\epsilon}\right)^{2s+3},
\end{eqnarray}
\\where $8.74\times 10^{-7}\leq \tilde{\sigma}\leq 1$ (lower and upper limits 
referring to $s=3/2$ and -3/2 respectively), and the dimensionless parameter 
$\sigma_0$ is given by:

\begin{eqnarray}
   \sigma_0 = \frac{L_\mathrm{bol}}{12c^2\Sigma_0 R_\mathrm{g}^2},
\end{eqnarray}
\\where $L_\mathrm{bol}$ is the bolometric luminosity produced by 
the accretion process onto the compact object.

   As we have noted in Section 2, the disk has to be optically thick 
in order to be unstable to warping by irradiation. This implies the 
existence of an independent lower limit for the surface density, represented 
by $\Sigma_0^{\tau}$, which is not necessarily more restrictive than 
$\Sigma_0^{v_\mathrm{r}}$.

  Note also that $\Sigma_0^{\tau}$ depends on which mechanism is the 
main contributor to the disk opacity; if Thomson scattering on 
free electrons of cross-section $\sigma_\mathrm{T}$ dominates over 
free-free absorption (with opacity $\kappa_\mathrm{ff}$), the optical 
depth can be calculated from $\tau_\mathrm{T} = \sqrt{\sigma_\mathrm{T}\kappa_\mathrm{ff}/m_\mathrm{H}}\Sigma$, 
otherwise $\tau_\mathrm{ff} = \kappa_\mathrm{ff}\Sigma$ \citep{shsu73}.

   The Rosseland mean opacity for free-free absorption is 
$\kappa_\mathrm{ff}=\kappa_0\rho T^{-7/2}$, where $\rho$ and $T$ are 
the mass density and temperature of the accretion disk respectively, 
while $\kappa_0$ is a numerical constant equal to 
$\sim 3\times 10^{23}$ cm$^2$ g$^{-1}$ (e.g., \citealt{mal98}). Expressing $\rho$ 
and $T$ in terms of $\Sigma$ and $c_\mathrm{s}$, and assuming a non-relativistic 
ideal equation-of-state for the gas in the disk, $\kappa_\mathrm{ff}$ is 
given as:

\begin{eqnarray}
   \kappa_\mathrm{ff}(r) = \kappa_0\left(\frac{5}{3}\frac{k_\mathrm{B}}{\mu m_\mathrm{H}}\right)^{7/2}\frac{\Sigma(r)\Omega(r)}{c_\mathrm{s}^8(r)},
\end{eqnarray}
\\where $k_\mathrm{B}$ is Boltzmann's constant, $\mu$ is the mean 
molecular weight and $m_\mathrm{H}$ is the atomic Hydrogen mass.

   If $\kappa_\mathrm{ff}<\sigma_\mathrm{T}/m_\mathrm{H}$ at 
the critical radius, the lower limit on $\Sigma_0$ (in order 
to have an optically thick disk) is:

\begin{eqnarray}
   \Sigma_0^{\tau} > \left(\frac{m_\mathrm{H}}{\sigma_\mathrm{T}}\right)^{1/7}K_\mathrm{\tau}^{1/7}\left(\frac{c}{R_\mathrm{g}}\right)^{3/7}\left(\frac{\dot{M}}{\alpha}\right)^{4/7}\left(\frac{R_\mathrm{cr}}{R_\mathrm{g}}\right)^{-(s+9/14)}G_\mathrm{\tau},
\end{eqnarray}
\\where

\begin{eqnarray}
   K_\mathrm{\tau} = \left(\frac{5}{6\pi}\right)^4\left(\frac{5}{3}\frac{k_\mathrm{B}}{\mu m_\mathrm{H}}\right)^{-7/2}\kappa_0^{-1},
\end{eqnarray}
\\and

\begin{eqnarray}
   G_\mathrm{\tau}(a_\ast,R_\mathrm{cr}) = \left[1+a_\ast\left(\frac{R_\mathrm{cr}}{R_\mathrm{g}}\right)^{-3/2}\right]^{-1}F_\mathrm{Q}^{-4/7}(a_\ast,R_\mathrm{cr}).
\end{eqnarray}

   On the other hand, if free-free absorption dominates:

\begin{eqnarray}
   \Sigma_0^{\tau} > K_\mathrm{\tau}^{1/6}\left(\frac{c}{R_\mathrm{g}}\right)^{1/2}\left(\frac{\dot{M}}{\alpha}\right)^{2/3}\left(\frac{R_\mathrm{cr}}{R_\mathrm{g}}\right)^{-(s+3/4)}G_\mathrm{\tau}^{7/6},
\end{eqnarray}

   Thus, the lower limit for $\Sigma_0$ in the case of irradiation 
torques is obtained from the more restrictive value between 
$\Sigma_0^{v_\mathrm{r}}$ (equation 11) and $\Sigma_0^{\tau}$.

\subsection{Magnetically-driven instability}

   The influence of magnetic fields on accretion disks has been studied by 
several authors, especially in the cases in which the central object is magnetized (e.g., 
\citealt{aly80,lish80,lai99,tepa00}). The production of quasi-periodic oscillations
and luminosity variability in such objects has also been explored in the 
framework of external magnetic fields and accretion disk interactions (e.g., 
\citealt{aga97,tepa00,shla02a,shla02b}).

   \citet{lish80} and \citet{lai99} showed that when the spin axis of the magnetized 
object, as well as its dipole moment, are not aligned with the angular momentum 
of the disk, magnetic torques will be generated, causing warping and disk precession. 
Extending those results to non-magnetized central sources, \citet{lai03} showed 
that a large-scale magnetic field (associated with magnetically driven outflows) 
threading the disk also induces a warping instability and a retrograde precession. 
Nonlinear simulations of warped, viscous accretion disks driven by magnetic 
torques were performed by \citet{pfla04}. These numerical simulations showed 
that the disk can keep a steady-state warped shape with a rigid-body precession.

   In order to examine the applicability of the magnetically driven mechanism to 
our sample of objects, we have considered a very idealized configuration, similar 
to that adopted by \citet{lai03}. Let us assume that the accretion disk is threaded 
by a poloidal magnetic field $\vec{B_\mathrm{p}}$ with a radial pitch angle 
$\varphi = \arctan\vert B_\mathrm{R}/B_\mathrm{Z}\vert$, where $B_\mathrm{R}$ and 
$B_\mathrm{Z}$ are respectively the radial and parallel magnetic field components 
in relation to the direction of the compact object's rotation axis (Z-direction). 

   The poloidal magnetic field lines will be twisted by the disk
rotation, generating a toroidal component $B_\mathrm{\Phi}$ that has different signs 
above and below the disk plane due to the discontinuity of $B_\mathrm{R}$ at the 
disk mid-plane \citep{lai03}. Following \citet{lai03}, we introduce the azimuthal 
pitch $\zeta$, such that $B_\mathrm{\Phi} = \mp\zeta B_\mathrm{Z}$, with the negative and 
positive signs referring to $B_\mathrm{\Phi}$ above and below the disk mid-plane, 
respectively.

   In addition, the accretion disk plane does not have to be perpendicular to $B_\mathrm{Z}$, being 
instead tilted by an angle $\beta$ (see figure 2 in \citealt{lai03}). In this case, the projection 
of $B_\mathrm{Z}$ onto the perpendicular direction (to the disk) will be $B_\mathrm{Z}\cos\beta$, 
while the toroidal magnetic field will be $\mp\zeta B_\mathrm{Z}\cos\beta$.

   The discontinuities in the radial and toroidal components of the magnetic field inside 
the disk produce a net disk surface current. The interaction between the current's toroidal 
component and the radial component of the magnetic field leads to the magnetically driven precession, 
while the interaction between the toroidal component of the magnetic field and the radial 
component of the net disk surface current warps the disk. The magnetic torques per unit 
area responsible for the disk precession and warping (averaged over the azimuthal angle 
$\phi$) were calculated by \citet{lai03}:

\begin{eqnarray}
   \langle T_\mathrm{prec}^\mathrm{mag}\rangle_\phi = -\frac{\tan\varphi}{4\pi}r B_\mathrm{Z}^2,
\end{eqnarray}

\begin{eqnarray}
   \langle T_\mathrm{warp}^\mathrm{mag}\rangle_\phi = -\frac{\zeta}{4\pi}r B_\mathrm{Z}^2\cos\beta.
\end{eqnarray}

   We can see that this mechanism produces retrograde precession, as in the case of 
the tidal torques in binary systems, and it pulls the normal to the disk plane away 
from the $B_\mathrm{Z}$-direction, increasing the angle $\beta$ with time.

   The precession period induced by the magnetic torques can be estimated by: 

\begin{eqnarray}
    P_\mathrm{prec}^\mathrm{mag} = \frac{J_\mathrm{d}(R_\mathrm{ms},R_\mathrm{prec})}{\int^{R_\mathrm{prec}}_{R_\mathrm{ms}}\langle T_\mathrm{prec}^\mathrm{mag}\rangle_\phi rdr},
\end{eqnarray}
\\where $J_\mathrm{d}$ is the integrated angular momentum of the accretion disk:

\begin{eqnarray}
    J_\mathrm{d}(R_\mathrm{ms},R_\mathrm{prec}) = 2\pi\int^{R_\mathrm{prec}}_{R_\mathrm{ms}}L_\mathrm{d}(r)rdr.
\end{eqnarray}

   The magnetic warping timescale can be calculated using:

\begin{eqnarray}
    P_\mathrm{warp}^\mathrm{mag} = \frac{J_\mathrm{d}(R_\mathrm{ms},R_\mathrm{prec})}{\int^{R_\mathrm{prec}}_{R_\mathrm{ms}}\langle T_\mathrm{warp}^\mathrm{mag}\rangle_\phi rdr}.
\end{eqnarray}

   To calculate the precession timescale, we have assumed a 
power-law profile for $B_\mathrm{Z}$, as in \citet{lai98}, 
such that:

\begin{eqnarray}
  B_\mathrm{Z}(r) = B_{0,\mathrm{Z}}\left(\frac{r}{R_\mathrm{g}}\right)^\chi,
\end{eqnarray}
\\where $\chi=0$ correspond to the case of a constant $B_\mathrm{Z}$ along the disk, 
while $\chi=-3$ corresponds to the dipole case.

   Substituting our power-law parameterizations for $\Sigma$ and $B_\mathrm{Z}$ 
into equation (25), we can calculate the relation between the observationally 
inferred precession period and $\Sigma_0$ and $B_\mathrm{0,Z}$, expressed by the 
ratio $\tan\varphi P_\mathrm{prec}^\mathrm{mag}B_\mathrm{0,Z}^2/\Sigma_0$, as 
a function of the geometrical parameters $a_\ast$, $R_\mathrm{out}$, $s$ and 
$\chi$.

\subsection{Frame dragging and disk viscosity: the Bardeen-Petterson effect}

   Frame dragging produced by a rotating compact body with angular momentum 
$J_\mathrm{B}$ causes precession of a particle if its orbital plane is inclined 
in relation to the equatorial plane of the rotating object. This is known as the 
Lense-Thirring effect \citep{leth18}. The precession angular velocity $\Omega_{\mathrm{LT}}$ 
produced by the frame dragging is given by (e.g., \citealt{wilk72}):

   \begin{eqnarray}
      \Omega_\mathrm{LT}(r) = \frac{2G}{c^2}\frac{J_\mathrm{B}}{r^3}.
   \end{eqnarray}

   The presence of the Lense-Thirring effect in astrophysical systems with 
rotating neutron stars or Kerr black holes has been claimed in several works in 
the literature, as the physical mechanism behind the observed quasi-periodic 
oscillations (e.g., \citealt{cui98,stvi98,mala98}).

   The combined action of the Lense-Thirring effect and the internal viscosity 
of the accretion disk forces the alignment between the angular momenta of 
the Kerr black hole and the accretion disk. This is known as the Bardeen-Petterson 
effect \citep{bape75}, and it tends to affect only the innermost part of the 
disk due to the short range of the Lense-Thirring effect, while the disk's 
outer part tends to remain in its original configuration. The transition 
radius between these two regimes is known as the Bardeen-Petterson radius, 
$R_\mathrm{BP}$, and its location depends mainly on the physical properties 
of the accretion disk (\citealt{bape75,kupr85,ivil97,nepa00}).

   A rough estimate of $R_\mathrm{BP}$ can be obtained by comparing the time-scales 
for Lense-Thirring precession and warp transmission through the disk (e.g., \citealt{naar99}). 
If the transmission occurs diffusively, the Bardeen-Petterson radius can be obtained 
from: 

\begin{eqnarray}
   R_\mathrm{BP}^\mathrm{diff}=\sqrt{\frac{\nu_2(r=R_\mathrm{BP}^\mathrm{diff})}
     {\Omega_\mathrm{LT}(r=R_\mathrm{BP}^\mathrm{diff})}}.
\end{eqnarray} 

    The diffusive regime in a Bardeen-Petterson disk has been explored by several 
authors (e.g., \citealt{kupr85,scfe96,nepa00,fran05,lopr06}), either using analytical 
calculations or numerical methods.  

   A Bardeen-Petterson disk in a wave-like regime has also been studied in the 
literature (e.g., \citealt{ivil97,ogil99,lub02}); in such a situation, 
$R_\mathrm{BP}$ is given by:

\begin{eqnarray}  
   R_\mathrm{BP}^\mathrm{w} = \frac{c_\mathrm{s}(r=R_\mathrm{BP}^\mathrm{w})}
      {\Omega_\mathrm{LT}(r=R_\mathrm{BP}^\mathrm{w})}.
\end{eqnarray}

   The transition between the diffusive and wave-like regimes occurs 
approximately at a radius $R_\mathrm{T}\sim H/\alpha$ \citep{pali95}.
 
   The time-scale for the black hole to align its angular momentum with that 
of the accretion disk was first estimated by \citet{rees78}. \citet{scfe96} 
obtained an analytic solution to the equations that control the warp evolution 
in the case of a disk with constant surface density and used that to calculate 
the alignment time-scale. \citet{naar99} generalized the results found by 
\cite{scfe96} to a power-law viscosity. These studies suggest that the alignment 
time-scale can be estimated by:

\begin{eqnarray}
   T_\mathrm{align} =
      J_\mathrm{B}\left(\frac{dJ_\mathrm{B}}{dt}\right)^{-1}\sin\varphi,
\end{eqnarray}
\\where $\varphi$ is the angle between the angular momentum of the 
neutron star/black hole $J_\mathrm{B}$ and the direction perpendicular 
to the outer disk, and the time derivative of $J_\mathrm{B}$ is given by:

\begin{eqnarray}
   \frac{dJ_\mathrm{B}}{dt} =
       -2\pi\sin\varphi\int_{R_\mathrm{BP}}^{R_\mathrm{out}}\Omega_\mathrm{LT}(r)L_\mathrm{d}(r)rdr.
\end{eqnarray}

   Contrary to \citet{scfe96}, \citet{kin05} showed that counter-alignment 
can occur when the spins are anti-parallel ($\varphi>\pi/2$) and the angular 
momentum on the disk is smaller than about twice that of the rotating 
compact object ($J_\mathrm{d}<2J_\mathrm{B}$). The results obtained recently 
by \citet{lopr06} support the possible existence of counter-alignment.

   Other than in the special circumstances mentioned above, the 
Bardeen-Petterson effect forces the disk to align gradually with 
the rotating compact object. According to \citet{scfe96}, the time 
evolution of the alignment, and the precession period of the disk 
angular momentum around the spin axis of the rotating accreting object, 
$P_\mathrm{prec}^\mathrm{BP}$, are given by:

\begin{equation}
   \varphi(t) = \varphi_0e^{-\Delta t/T_\mathrm{align}}
\end{equation}

\begin{equation}
   P_\mathrm{prec}^\mathrm{BP}(t) = P_\mathrm{0}e^{-\Delta t/T_\mathrm{align}}
\end{equation}
\\where $\Delta t = t-t_0$, and $\varphi_0$ and $P_\mathrm{0}$ are, respectively, 
the inclination angle and precession period at time $t_0$, when the action of the 
Bardeen-Petterson torques over the accretion disk started ($t_0\leq 0$ is measured 
in the past from the present time). \citet{scfe96} found that the timescales for 
precession and realignment are identical, implying that $P_0=T_\mathrm{align}$, 
which will also be used here.

\section{Source sample}

   To analyze separately the consequences of each of the precession mechanisms 
discussed in the last section, we selected eight X-ray binaries and four AGNs that present 
signatures of warping and/or precession in their accretion disks. In 
this section we introduce these systems and provide their basic 
parameters (the ones that will be used in our calculations).

   X-ray binaries, characterized as semi-detached systems with one 
component filling its critical Roche lobe, can 
be roughly divided into two categories: high- and low-mass X-ray binaries 
(HMXBs and LMXBs respectively).

   HMXBs are composed of an accreting compact source (a black hole or 
a neutron star) and either an OB supergiant or a Be star \citep{par83}, 
which implies that such systems are short-lived, with ages of less than 
about $2\times 10^7$ yr \citep{heu94}. They are spatially distributed 
in the galactic plane, being associated with young stellar populations.

   By contrast, LMXBs contain a neutron star accreting from a low-mass 
star ($\lesssim 2 M_\sun$). They are typically associated with an older stellar 
population, with ages of about one Gyr \citep{heu94}.

   Here, we will use eight X-ray binaries with regular (or 
quasi-regular) precession periods inferred basically from the optical 
and X-ray continuum variability, as well as from variability in the line 
intensity and velocity in some cases. Our sample is composed of two 
LMXBs (Her\,X-1 and Cyg\,X-2) and six HMXBs, four of which have an accreting 
neutron star (LMC\,X-4, Cen\,X-3, SMC\,X-1 and 4U\,1907+09) while the 
remaining sources are black hole systems (SS\,433 and Cyg\,X-1). It is 
important to note that SS\,433 is one of the best studied microquasars 
in the literature, with a prominent precessing jet/counterjet (e.g., 
\citealt{hjjo81,marg84,blu01,blbo04}). The relevant parameters of the sources, 
for the present work, are listed in Table 1. In particular, the 
values of the magnetic field strength listed in Table 1 refer basically 
to the surface magnetic fields of neutron stars, while for the two black 
hole systems we used values derived from dynamo models for the disk 
magnetic field \citep{pufa82,rose95}.

   Some AGNs also exhibit signatures of warping/precession in their 
accretion disks. The signatures are in the form variability of 
double-peaked Balmer lines and of the continuum emission, or distortions 
in the jet morphology (variations in the jet orientation and jet velocity) 
(e.g., \citealt{abra00,sto03,caab04b}). We have selected four 
AGNs with these characteristics: the Seyfert 1 galaxy 3C\,120, 
the Seyfert 2 galaxy NGC\,1068, the broad-line radio galaxy Arp\,102B, 
and the BL Lac OJ\,287. Their parameters are listed in Table 2. 
In relation to their magnetic field strengths, since there is 
evidence in some cases that the strength of the magnetic field 
is not in equipartition (e.g. \citealt{come00}), we have chosen 
the field strength to be equal to that which has been observationally 
determined in a few AGN ($\sim 10^4$ G; e.g., \citealt{firo93,loba98}).

\section{Analysis}

   We have analyzed which of the four mechanisms can be driving the 
observed precession in the twelve sources of our sample. Below we 
describe the results of this analysis, discussing each mechanism 
separately.

\subsection{Tidal torques in a binary system}

   In order to examine the possibility that tidal torques induced 
by the companion affect the accretion disk, we have plotted in the 
left panel of Figure 1 the quantity $\cos\theta P_\mathrm{prec}/P_\mathrm{orb}$ 
predicted by equation (14) as a function of $q$ (thick line); 
the values for all the X-ray binaries in our sample are also 
displayed (star symbols). The values of $\theta$ for Her\,X-1, 
LMC\,X-4, SS\,433 and Cyg\,X-1 were taken from the literature 
\citep{heva89,sco00,sti02,rom02}, while $\theta=10\degr$ was 
assumed for the other X-ray binaries.

   The gray regions in Figure 1 show the part of parameter space 
where tidal torques from a companion can drive precession. Clearly, 
the observed ratios between the precession and orbital periods for Cen\,X-3 
and Cyg\,X-2 reside below of this region. Therefore, the origin of 
their disk precession cannot be attributed to the tidal torques of 
the secondary. In the case of Cyg\,X-1, \citet{lac06} have recently 
claimed that its observed precession is prograde, which would also 
rule out the tidal torque scenario. For the other five X-ray binaries, 
precession may be induced by the torques of the companion star. Although 
\citet{katz73} and \citet{larw98} have already proposed that tidal 
torques could lead to the observed precession in a few of these sources, 
tidally-induced precession has not been previously suggested (to the 
best of our knowledge) for 4U\,1907+09.

   In the case of the AGN sources, there is no straightforward 
observation that supports the existence of a supermassive binary 
black hole system in any of their nuclear regions. Nevertheless, the continuum 
variability and the anomalous jet kinematics in OJ\,287 and 3C\,120 
have been used to put some constraints on the physical parameters 
of a possible binary system \citep{sil88,abra00,caab04b}. In the 
right-hand panel of Figure 1, we show the upper limit for 
$\cos\theta P_\mathrm{prec}/P_\mathrm{orb}$ as a function of $q$ for 
our extragalactic sources, considering a timescale for losses 
due to gravitational radiation $\tau_\mathrm{GW}$ \citep{shte83} 
of 1000 yr. Although this value has been chosen arbitrarily, it 
guarantees that no significant changes in the orbit of the 
secondary occur on such a timescale, so that the observed 
precession periods do not vary substantially. Note that increasing 
(decreasing) $\tau_\mathrm{GW}$ by a factor of 10 results 
in a decrease (increase) in the upper limit for $\cos\theta P_\mathrm{prec}/P_\mathrm{orb}$ 
only by a factor of $\sim$2.4. For OJ\,287, 3C\,120 and NGC\,1068, 
$\theta$ was taken from the literature \citep{abra00,caab04b,cap06}, 
while for Arp\,102B, we assumed $\theta=10\degr$.

   By contrast with the galactic binaries, we do not expect $q\gtrsim 1$ 
in supermassive binary black hole systems, since the expectation is generally 
that the accretion disk responsible for the AGN activity is associated 
with the more massive black hole. Except for Arp\,102B, it is always 
possible to find a value of $q\leq1$ for which $\cos\theta P_\mathrm{prec}/P_\mathrm{orb}$ 
is within the gray area. In addition, the separations between the putative 
black holes in NGC\,1068 obtained from our calculations
\footnote{Typically 100-10$^4 R_\mathrm{g}$, depending on the timescales 
for losses due to gravitational radiation.} 
are always smaller than the dimensions of the maser disk. Consequently, 
we would expect a more complex disk morphology than that suggested by 
the observations. Thus, tidal forces of an orbiting black hole can (in 
principle at least) induce the inferred precession rate in 3C\,120 and 
OJ\,287 (and perhaps NGC\,1068).

\subsection{Irradiation-driven torques}

   In Figures 2 and 3 we present the precession period 
induced by the radiation torques as a function of the 
critical radius at which the disk becomes irradiation-warping 
unstable, for each source of our sample, using equations 
(15) and (16). Following \citet{mal98}, the calculations 
were performed for power-law disks, using $s=-3/2, -1, 0$ 
and 3/2. The lower and upper limits for each solution 
refer to the respective lower and upper limits of $\Sigma_0$ 
calculated from equations (9), (11) and (19), as discussed 
previously.

   The critical radius, as well as the precession period, 
depend on the ratio between $\eta$ and $\epsilon$, which 
is usually unknown. We have assumed $0.0377\leq\epsilon\leq 0.42$
\footnote{The upper and lower limits correspond to the 
accretion efficiencies from the maximum prograde and 
retrograde spinning black holes, respectively.}, while 
$1\leq\eta\leq f(\alpha)$, where $f(\alpha)=2(1+7\alpha^2)/[\alpha^2(4+\alpha^2)]$ 
\citep{ogil99}. In the case of $\eta=1$ (open symbols 
in Figures 2 and 3), $2.38\leq\eta/\epsilon\leq 26.49$, 
implying $223.80\leq R_\mathrm{cr}/R_\mathrm{g}\leq 2.77\times 10^4$ for $s=-3/2$ 
and $1.34\times 10^3\leq R_\mathrm{cr}/R_\mathrm{g}\leq 1.66\times 10^5$ 
for $s=3/2$. For $\eta\ne 1$, $\eta/\epsilon$, and 
consequently $P_\mathrm{prec}^\mathrm{rad}$ and $R_\mathrm{cr}$, 
also depend on the value of $\alpha$; we have chosen 
$\alpha=0.1$ for the X-ray binaries and $\alpha=0.01$ for the AGNs, 
except for NGC\,1068 for which $0.001\lesssim\alpha\lesssim 0.012$ 
\citep{cap06}. These values of $\alpha$ imply that the critical 
radii, for a fixed $\epsilon$, are systematically larger in the 
case of $\eta\ne 1$.

   For the X-ray binaries, in the case of an isothermal disk 
($s=-3/2$) with $\eta=1$ (open circles), the predicted precession 
period intervals cross the horizontal line, which represents the 
observed precession period. This indicates that the radiation instability 
can be responsible for precession in those X-ray binaries (for this 
particular accretion disk model). However, the situation changes 
when we take into account other combinations among $s$, $\eta$ 
and $\epsilon$; even though the predictions assuming $s=-3/2$ and 
$\eta\ne 1$ are compatible with the observed precession in most cases, 
power-law disks with $s=3/2$ generally fail to reproduce the observations, 
providing incompatible precession periods and/or critical radii larger 
than the outer disk radii.

   As pointed out by \citet{mal98}, disk precession induced by radiation 
must be prograde in the absence of external torques, which would be in 
contradiction with the observations of SS\,433, Her\,X-1 and LMC\,X-4. 
However, as shown by \citet{mabe97} and corroborated by \citet{mal98}, 
the quadrupole torque from a companion star might allow for the existence 
of prograde and retrograde precession modes. Thus, retrograde precession 
due to the irradiation mechanism is possible only if there is some 
additional torque acting upon the accretion disk.

   In the case of AGNs, the situation is quite different. For $\eta\ne 1$, 
the critical radius is larger than the outer disk radius for all AGNs in 
our sample, suggesting that in these cases the disk is stable against 
radiation torques.

   Thus, the irradiation instability can be responsible for the precession 
in our sample of X-ray binaries, in agreement with \citet{ogdu01}, but it 
is not favored in the case of the AGNs we have considered.

\subsection{Magnetically-driven torques}

   To analyze the precession induced by magnetic torques, we created a 
two-dimensional grid of the quantity $\tan\varphi P_\mathrm{prec}^\mathrm{mag}B_\mathrm{0,Z}^2/\Sigma_0$, 
fixing $s$ and $\chi$, as a function of $a_\ast$ and $R_\mathrm{out}$ 
for each source in our sample.

   These grids are plotted in Figures (4) and (5) for the X-ray binaries 
and in Figure (6) for the AGNs assuming power-law disks with $s=-2, -1$ and 0 
and power-law magnetic field configurations with $\chi=0, -1, -2$ and -3. The 
gray area represents the allowed range for 
$P_\mathrm{prec}^\mathrm{mag}B_\mathrm{0,Z}^2/\Sigma_0$ (assuming $\tan\varphi=1$) 
using the available constraints obtained from the observations. As in 
the case of tidal torques in binary systems, acceptable solutions 
must also obey $R_\mathrm{prec}\leq R_\mathrm{out}$. The dimensionless 
spin parameter influences only the location of the inner radius of the 
accretion disk, while the estimated extreme values for $\Sigma_0$ are 
responsible for the limits of the gray region in the panels of the 
Figures (4)-(6). 
  
   We can see in Figures (4)-(6) that not all magnetic field configurations 
provide solutions compatible with the observations. Nevertheless, it is 
possible to find combinations of $s$ and $\chi$ that reproduce the observed 
precession periods, which means that magnetically-driven instabilities cannot 
be excluded as a potential mechanism for the observed precession.

   At least in the case of the X-ray binaries with an accreting neutron star, 
such as Her\,X-1, LMC\,X-4, Cen\,X-3, Cyg\,X-2, SMC\,X-1 and 4U\,1907+09, 
the magnetic field configuration is dipolar to a good approximation 
(e.g., \citealt{aly80}). We have not found any possibility that disk 
precession in those systems can be associated with magnetic field configurations 
with $\chi\geq -1$, except for Cyg\,X-2 for which acceptable solutions are 
obtained with $\chi=-1$ and $\chi=0$ only if $s=-2$. For the black hole 
X-ray binaries, Cyg\,X-1 has similar results to those of Cyg\,X-2, while 
for the microquasar SS\,433 lower values of $\chi$ are strongly favored.

   In contrast to the neutron star X-ray binaries, there is no preferred 
magnetic field configuration in the case of AGNs. Our results show 
that the range of the allowed solutions is systematically narrower than 
that obtained for the X-ray binaries. In addition, it seems to have two 
different regimes in our AGN sample: if the accretion disk surface density 
decreases faster with radial distance, with $s\lesssim -1$, the z-component 
of the magnetic field must be radially constant in order to reproduce the 
observed precession periods; otherwise, it is necessary for the surface 
density not to vary radially ($s=0$) if the magnetic field decreases along 
the disk ($\chi\leq -1$).

\subsection{Bardeen-Petterson effect}

   In order to study the consequences of the Bardeen-Petterson 
effect in our sample of sources, we have separated the black hole systems 
from those with accreting neutron stars. In the former case, 
we have calculated the Bardeen-Petterson radius and the alignment timescale 
for six values of the black hole spin ($a_\ast=\pm 0.1, \pm 0.5$ 
and $\pm 1$) and for $s=-2, -1$ and 0, as shown in Figure 7. As 
in the previous sections, calculations were performed for 
$\eta=f(\alpha)$ ($\alpha=0.1$ and 0.01 for X-ray binaries and 
AGNs, respectively) and $\eta=1$, as well as for the estimated lower 
and upper limits on the accretion disk surface density.

   Once $s$ and $\eta$ are fixed, there are generally two values 
of $R_\mathrm{BP}$ that are determined by the lower and upper 
limits of $\Sigma_0$. In some cases the lower limit for $\Sigma_0$ 
leads to $R_\mathrm{BP}<R_\mathrm{ms}$, and because of that 
only the upper limit is shown in the figure. All sources have 
their Bardeen-Petterson radius inside of the outer edge of the 
disk, independently of the specific values for $s$, $\eta$ and 
$\Sigma_0$, which indicates that the Bardeen-Petterson mechanism 
is at least applicable.

   We also plot in Figure 7 the time interval necessary for 
each system to reach the observationally inferred precession period 
(using equation 35). As in the case of the alignment timescales, 
the plots were truncated at 13.7 Gyr, the age of the Universe as 
inferred from the WMAP results \citep{ben03}. Unfortunately, 
only in the case of SS\,433 we can constrain better the allowed 
model parameters using its estimated age (between $2\times10^4$ 
and $2\times10^5$ yr; \citet{zea80}). This age constraint clearly 
favors model disks with $\eta\ne 1$ at least for $|a_\ast|\leq 0.1$. 
The same conclusions are valid for Cyg\,X-1 if we take into account 
the upper limit for the age of HMXBs \citep{heu94}.

   In the case of NGC\,1068, we also have included the solutions 
obtained from a power-law accretion disk with $s=-1.05$ \citep{hure02}, 
corresponding to model A shown in Figure 2 of \citet{cap06}. 
Indeed, these authors showed that the Bardeen-Petterson effect 
can reproduce the disk configuration suggested by the maser 
observations \citep{gal04}, as well as the general shape of the 
parsec and kiloparsec radio jet. 

   For the neutron star systems in our sample, the spin period 
of the accreting neutron stars, $P_\mathrm{s}$, are known, and 
are listed in Table 3. From the definition of the angular 
momentum parameter, we can obtain (e.g., \citealt{stvi98}):
 
\begin{equation}
   a_\ast = \frac{2\pi c}{G}\frac{I_\mathrm{p}}{M_\mathrm{p}}\frac{\nu_\mathrm{s}}{M_\mathrm{p}}
\end{equation}
\\where $I_\mathrm{p}$ and $\nu_\mathrm{s}$ are respectively 
the moment of inertia and the spin frequency of the neutron star.

   If we write $I_\mathrm{p}=10^{45} I_\mathrm{45}$ g cm$^{2}$ 
and $M_\mathrm{p}=M_\mathrm{o}$ M$_\sun$ typically, 
$0.5<I_\mathrm{45}/M_\mathrm{o}<2$, considering 
rotating neutron star models for different equations of state 
and masses. In Table 3 we give the allowed ranges of $a_\ast$ 
for our neutron star systems using their respective orbital periods 
and the lower and upper limits for $I_\mathrm{45}/M_\mathrm{o}$.

   In Figure 8, we plot the results for the neutron star system 
displaying, as in the Figure 7, the Bardeen-Petterson radius, 
the alignment timescale and the time interval necessary for each 
system reach the observationally inferred precession period for 
the maximum absolute value of $a_\ast$ listed in Table 3. 
As in the case of the Kerr black holes, all sources have their 
respective Bardeen-Petterson radii inside the outer radius 
of the accretion disk. All spinning neutron stars in our sample 
reach the observed precession periods in a timescale shorter than 
20 Myr. In the case of Her\,X-1, whose age has been estimated 
previously \citep{ver90}, any allowed combination of model 
parameters leads to Her\,X-1 reaching its observed precession 
period via the Bardeen-Petterson effect. The situation is similar, 
even though somewhat more restrictive, for Cyg\,X-2.

\section{Conclusions}

   In this work, we have selected eight X-ray binaries and four 
AGNs that present signatures of precession in their accretion 
disks in order to examine the compatibility between their 
precession periods and the predictions from four distinct 
warping/precession physical mechanisms: tidal torques from 
a companion in a binary system, irradiation- and magnetically-driven 
instabilities, and the Bardeen-Petterson relativistic effect. 

   We have assumed a power-law surface density distribution 
for the accretion disks, constraining their physical parameters 
from observational data available in the literature.

   For the X-ray binaries in our sample, we found that tidal 
torques from a companion in a binary system provide precession 
timescales compatible with those inferred in SS\,433, Her\,X-1, 
LMC\,X-4 and SMC\,X-1, as indeed has been previously suggested 
in the literature \citep{katz73,larw98}. In addition, we showed 
that tidal torques can also drive precession in 4U\,1907+09, 
which (as far as we know) had not been proposed before this work. 
The mechanism can be ruled out for Cen\,X-3 and Cyg\,X-2. In the 
case of our AGN sample (assuming that they contain binary black 
holes), we have shown that tidal torques cannot produce the disk 
precession inferred in Arp\,102B.

   Although the irradiation-driven instability usually provides 
precession timescales that agree with those observed in our 
X-ray binary sample, it is unsuccessful in reproducing the 
reported precession periods for the four AGNs considered here. 
The critical radii at which those AGN accretion disks become 
unstable against radiation torques were found to be larger than 
the expected disks' outer radii (for $\eta\ne1$). Since the 
irradiation-driven instability produces prograde disk precession, 
the retrograde precession observed in SS\,433, Her\,X-1 and 
LMC\,X-4 may only be generated in the presence of some external 
torques. 

   We have shown that torques due to a misaligned magnetic field 
(with respect to the perpendicular direction to an accretion disk) 
can also reproduce the observed precession periods of our sample. 
For the accreting neutron stars, a dipolar configuration must 
be a good approximation to their magnetic fields; in such a case 
(or for $\chi<-1$), the magnetic-driven torquing can induce 
precession in the inner parts of the accretion disks at the 
rate inferred observationally. For the black hole X-ray binaries, 
magnetic configurations with $\chi\geq -1$ are favored. Although 
the geometry of the magnetic field is not usually constrained 
by observations in AGNs, our results indicate the existence 
of two distinct regimes: for accretion disks with 
$s\lesssim -1$, the z-component of the magnetic field must be 
radially constant in order to reproduce the observed precession periods; 
otherwise, it is necessary to have a constant surface density ($s=0$) with 
a magnetic field weakening radially along the disk ($\chi\leq -1$). 
In addition, if the accretion disk of Cyg\,X-1 is actually precessing 
progradely \citep{lac06}, we can rule out magnetically-driven 
torques as the cause of its precession.

   The Bardeen-Petterson effect produces precession timescales 
compatible with those observed in all sources of our sample, 
considering that the alignment of the accretion disk evolves 
on a similar timescale (as in \citealt{scfe96}). The timescale for reaching 
the observed precession rate is usually shorter than the estimated 
lifetime of the sources. Considering the accretion disk parameters 
used in this work, we have found that all sources in our sample 
have Bardeen-Petterson radii smaller than the disk's outer radius.

   NGC\,1068, for which maser observations have been used to 
infer the physical properties of the accretion disk, may 
provide the strongest candidate for the Bardeen-Petterson 
effect. In this case, we have shown that the irradiation-driven 
instability is incompatible with the observations. Tidal 
torques are also highly unlikely (even if a binary black 
hole were to be present in this system) since it would most 
likely result in other changes to the disk morphology, which 
are not observed. Magnetically-driven torques would work 
only if a rather contrived field configuration exists ($B_\mathrm{Z}$ 
constant along the precessing part of the disk). Our conclusions 
give further support to the analysis of \citet{cap06}, 
who showed that the Bardeen-Petterson effect is consistent 
with the warping morphology suggested by \citet{gal04}.

   The Bardeen-Petterson precession is an important 
general-relativistic effect, and its unambiguous detection 
would be of great value. This work has shown, however, 
that given the many observational uncertainties that 
still exist (both in X-ray binaries and in AGNs), it 
is extremely difficult to rule out definitively other 
precession mechanisms.

   It is important to emphasize that although we have 
analyzed separately the four precession mechanisms in this 
work, they are not mutually exclusive. This means that more 
than one mechanism might be operating in a given system.


%% Included in this acknowledgments section are examples of the
%% AASTeX hypertext markup commands. Use \url without the optional [HREF]
%% argument when you want to print the url directly in the text. Otherwise,
%% use either \url or \anchor, with the HREF as the first argument and the
%% text to be printed in the second.

\acknowledgments

A.C. acknowledges financial support from FAPESP (Procs. 03/13882-0 and 05/56266-2), 
as well as the hospitality of the Space Telescope Science Institute, where this 
work was carried out.

\clearpage

\begin{deluxetable}{cccccccccccccc}
\tabletypesize{\scriptsize}
\rotate
\tablecaption{Physical parameters of the X-ray binary sources.\label{tbl-1}}
\tablewidth{0pt}
\tablehead{
\colhead{Source} & \colhead{$M_\mathrm{p}$ (M$_\sun$)} & \colhead{Ref.} 
& \colhead{$q$\tablenotemark{b}} & \colhead{$P_\mathrm{orb}$ (d)} & \colhead{Ref.} 
& \colhead{$P_\mathrm{prec}$ (d)} & \colhead{Ref.} & \colhead{Type\tablenotemark{d}} & \colhead{Ref.}
& \colhead{$B$ (G)} & \colhead{Ref.} & \colhead{$L_\mathrm{bol}$ (erg s$^{-1}$)\tablenotemark{f}} & \colhead{Ref.}
}
\startdata
SS\,433     &  11~$\pm$~3           & 1  &   1.73  &  13.08 & 8  &  162.5                & 15   & R & 23,24   &  10$^5$                            &  28  &  39.74                   &  36 \\
Her\,X-1    &  1.50~$\pm$~0.30      & 2  &   1.53  &  1.7   & 9  &  34.88                & 16   & R & 25      &  3.5$\times 10^{12}$               &  29  &  37.58                   &  2  \\
LMC\,X-4    &  1.38~$\pm$~0.25      & 3  &   10.6  &  1.408 & 3  &  30.275               & 17   & R & 26      &  10$^{13}$                         &  30  &  38.19                   &  37 \\
Cen\,X-3    &  1.21~$\pm$~0.21      & 4  &   16.9  &  2.087 & 10 &  140\tablenotemark{c} & 18   & ? & -       &  3.5$\times 10^{12}$               &  31  &  37.88                   &  38 \\
Cyg\,X-1    &  10.1                 & 5  &   1.76  &  5.67  & 11 &  142                  & 19   & P & 27      &  10$^8$\tablenotemark{e}           &  32  &  37.80                   &  39 \\
Cyg\,X-2    &  1.78~$\pm$~0.23      & 6  &   0.34  &  9.844 & 12 &  77.7                 & 20   & ? & -       &  4.5$\times 10^8$\tablenotemark{c} &  33  &  37.65                   &  40 \\
SMC\,X-1    &  1.60~$\pm$~0.10      & 7  &  10.75  &  3.893 & 13 &  60                   & 21   & ? & -       &  7.6$\times 10^{12}$               &  34  &  38.30                   &  41 \\
4U\,1907+09 &  1.4\tablenotemark{a} & -  &   21.4  &  8.375 & 14 &  41.6                 & 22   & ? & -       &  2.1$\times 10^{12}$               &  35  &  37.72\tablenotemark{g}  &  42 \\
\enddata

\tablerefs{
(1) \citealt{gie02}; (2) \citealt{rey97}; (3) \citealt{lev91}; (4) \citealt{ash99}; (5) \citealt{her95}; 
(6) \citealt{orku99}; (7) \citealt{rey93}; (8) \citealt{ste87}; (9) \citealt{tan72}; (10) \citealt{pau05}; 
(11) \citealt{pri95}; (12) \citealt{cas98}; (13) \citealt{sch72}; (14) \citealt{int98}; (15) \citealt{sti02};
(16) \citealt{gia73}; (17) \citealt{tslu05}; (18) \citealt{prte83}; (19) \citealt{bro99}; (20) \citealt{wij96};
(21) \citealt{woj98}; (22) \citealt{prte84}; (23) \citealt{leib84}; (24) \citealt{bri89}; (25) \citealt{gebo76}; 
(26) \citealt{heva89}; (27) \citealt{lac06}; (28) \citealt{rose95}; (29) \citealt{dal98}; (30) \citealt{lab01}; 
(31) \citealt{bur00}; (32) \citealt{pufa82}; (33) \citealt{camp00}; (34) \citealt{kali99}; (35) \citealt{cob02}; 
(36) \citealt{okud02}; (37) \citealt{vrt97}; (38) \citealt{labe98}; (39) \citealt{zdz02}; (40) \citealt{vrt90}; 
(41) \citealt{woj00}; (42) \citealt{muk01}.}

%%\tablecomments{The quoted errors in $M_\mathrm{BH}$ and $s$ are 
%%given by \citet{hure02} and \citet{lobe03}, while for $\Sigma_0$ 
%%were obtained from error propagation.}

\tablenotetext{a}{Mass value assumed in this work;}
\tablenotetext{b}{Quoted errors obtained from error propagation;}
\tablenotetext{c}{Mean value;}
\tablenotetext{d}{Type of precession (in relation to the angular momentum of the accretion disk: prograde (P), retrograde (R). The symbol "?" indicates 
                     that there is no available information;}
\tablenotetext{e}{Upper limit;}
\tablenotetext{f}{Base-10 logarithm of the bolometric luminosity;}
\tablenotetext{g}{Flare state (upper limit).}

\end{deluxetable}

\clearpage

\begin{deluxetable}{ccccccccc}
\tabletypesize{\scriptsize}
%%\rotate
\tablecaption{Physical parameters of the AGNs.\label{tbl-2}}
\tablewidth{0pt}
\tablehead{
\colhead{Source} & \colhead{$M_\mathrm{tot}$ (10$^8$ M$_\sun$)} & \colhead{Ref.}  
& \colhead{$P_\mathrm{prec}$ (yr)\tablenotemark{a}} & \colhead{Ref.} & \colhead{Type\tablenotemark{c}} 
& \colhead{$B$ (G)\tablenotemark{d}} & \colhead{$L_\mathrm{bol}$ (erg s$^{-1}$)\tablenotemark{e}} & \colhead{Ref.}
}
\startdata
NGC\,1068   &  0.120  & 1 &  (2.08-35.2)$\times 10^{3}$\tablenotemark{b}  & 5 &  ?  & 10$^4$   &  44.84\tablenotemark{f}  &  9  \\
3C\,120     &  0.556  & 2 &  11.9                                         & 6 &  ?  & 10$^4$   &  45.34                   &  10 \\
OJ\,287     &  6.170  & 3 &  8.88                                         & 7 &  ?  & 10$^4$   &  44.44\tablenotemark{g}  &  3  \\
Arp\,102B   &  1.380  & 4 &  2.15                                         & 8 &  ?  & 10$^4$   &  43.41                   &  4  \\
\enddata

\tablerefs{
(1) \citealt{hure02}; (2) \citealt{pet04}; (3) \citealt{wan04}; (4) \citealt{wuli04}; (5) \citealt{cap06}; 
(6) \citealt{caab04b}; (7) \citealt{abra00}; (8) \citealt{new97}; (9) \citealt{gal04}; (10) \citealt{wour02}.}

%%\tablecomments{The quoted errors in $M_\mathrm{BH}$ and $s$ are 
%%given by \citet{hure02} and \citet{lobe03}, while for $\Sigma_0$ 
%%were obtained from error propagation.}

\tablenotetext{a}{Values measured at the present time and in the source's reference frame;}
\tablenotetext{b}{Obtained from the jet kinematics and the Bardeen-Petterson model \citep{cap06};}
\tablenotetext{c}{Same nomenclature of the Table 1;}
\tablenotetext{d}{Adopted value for the magnetic field at $R_\mathrm{g}$;}
\tablenotetext{e}{Base-10 logarithm of the bolometric luminosity;}
\tablenotetext{f}{This value is a lower limit;}
\tablenotetext{g}{Assuming also that the bolometric luminosity is about ten times larger than the luminosity of the broad line region \citep{netz90}.}

\end{deluxetable}

\begin{deluxetable}{cccc}
\tabletypesize{\scriptsize}
%%\rotate
\tablecaption{Rotation period and spin parameter for the neutron stars in our X-ray binaries sample.\label{tbl-3}}
\tablewidth{0pt}
\tablehead{
\colhead{Source} & \colhead{$P_\mathrm{s}$ (s)} & \colhead{Ref.} & \colhead{$a_\ast$\tablenotemark{a}} 
}
\startdata
Her\,X-1    &  1.2377697~$\pm$~0.0000003  & 1  &  (1.92-7.69)$\times 10^{-4}$  \\
LMC\,X-4    &  13.509~$\pm$~0.002        & 2  &  (1.91-7.66)$\times 10^{-5}$  \\
Cen\,X-3    &  4.834477~$\pm$~0.000007    & 3  &  (0.61-2.44)$\times 10^{-4}$  \\
Cyg\,X-2    &  0.00606~$\pm$~0.00024      & 4  &  (0.33-1.32)$\times 10^{-1}$  \\
SMC\,X-1    &  0.706707~$\pm$~0.000001    & 5  &  (0.32-1.26)$\times 10^{-3}$  \\
4U\,1907+09 &  440.5738~$\pm$~0.0002      & 6  &  (0.58-2.31)$\times 10^{-6}$  \\
\enddata

\tablerefs{
(1) \citealt{oos01}; (2) \citealt{vrt97}; (3) \citealt{van80}; (4) \citealt{fock96}; (5) \citealt{kali99}; 
(6) \citealt{bay01}.}

%%\tablecomments{The quoted errors in $M_\mathrm{BH}$ and $s$ are 
%%given by \citet{hure02} and \citet{lobe03}, while for $\Sigma_0$ 
%%were obtained from error propagation.}

\tablenotetext{a}{The lower and upper limits refer respectively to $I_\mathrm{45}/M_\mathrm{o} = 0.5$ and 2 \citep{stvi98}.}

\end{deluxetable}

\clearpage

\begin{figure*}
%%\epsscale{.80}
\plotone{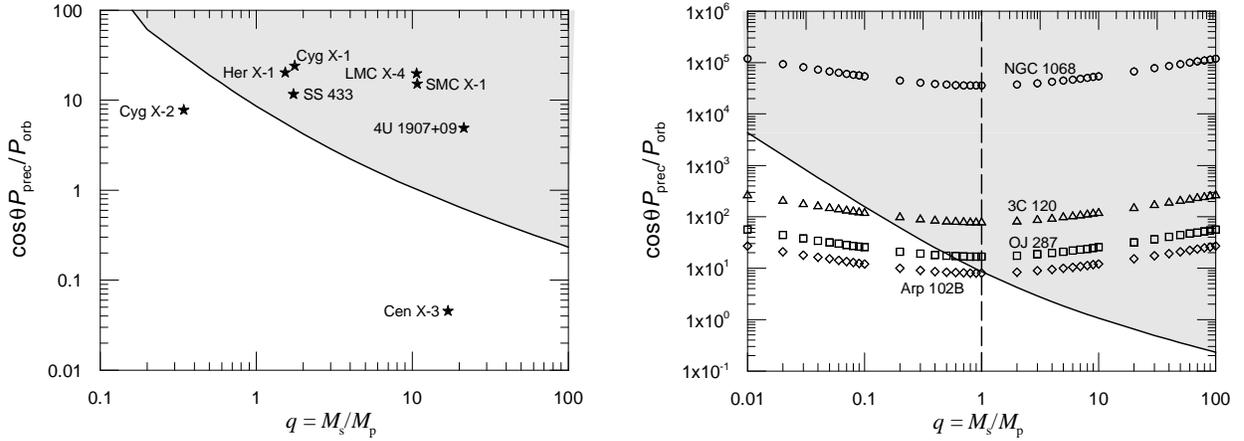}
\caption{Precession due to tidal torques induced by the companion in binary systems. 
{\it Left panel}: Ratio between the precession and orbital periods (multiplied by the 
cosine of the inclination angle of the orbital plane in relation to the 
disk plane) as a function of the mass ratio between the secondary and primary objects. 
The solid line is the theoretical prediction considering that the entire disk with an 
outer radius of 0.88$R_\mathrm{L}$ precesses rigidly. Data from our sample of X-ray 
binaries are displayed by the stars. Sources located in the gray area can have their 
disk precession driven by the companion's torque. {\it Right panel}: Upper limit for 
the AGNs of our sample considering a timescale for losses due to gravitational radiation 
equal to 1000 yr. The dashed vertical line marks $q=1$.}
\label{BinarySystem_Torques}
\end{figure*}

\clearpage

\begin{figure*}
\epsscale{.70}
\plotone{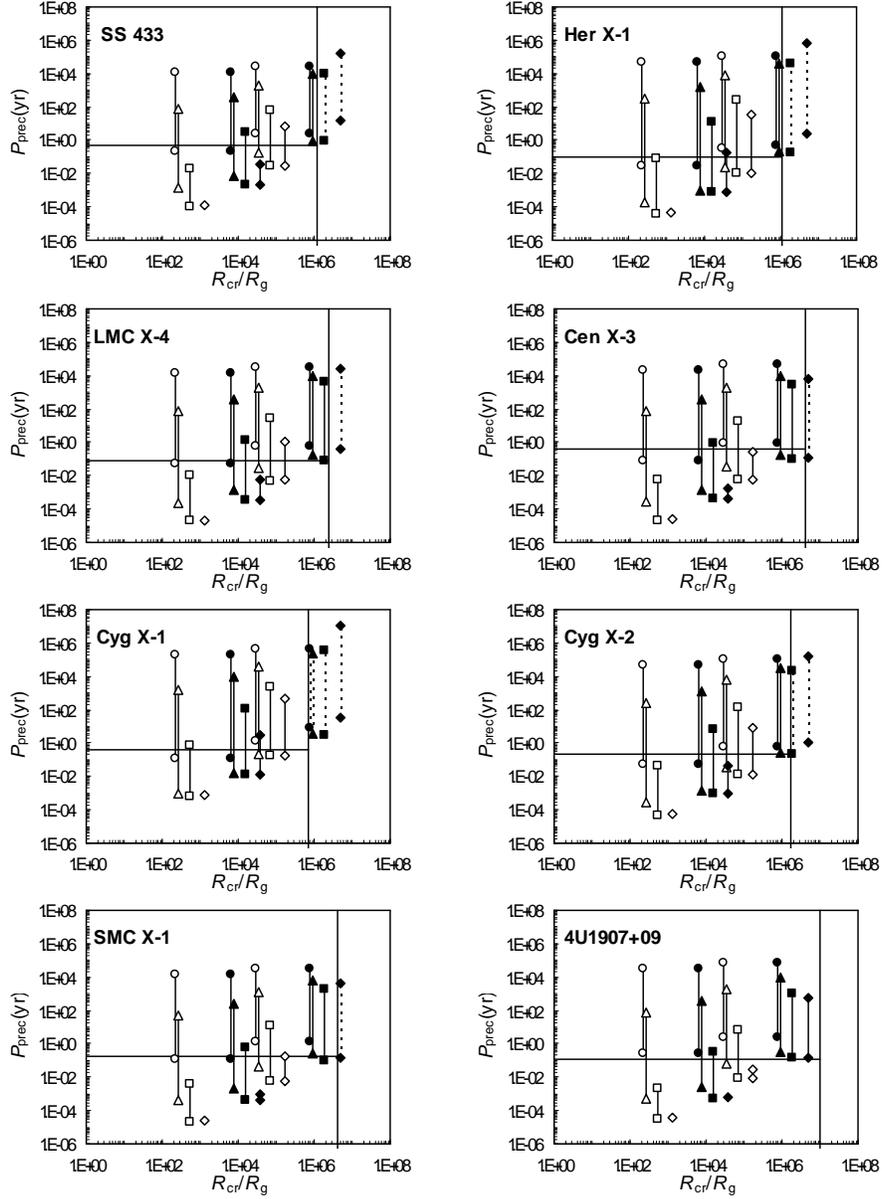}
\caption{Precession due to the radiation torques in our sample of X-ray binaries. Circles, 
triangles, squares and diamonds correspond respectively to $s=-3/2, -1, 0$ and 3/2. Filled symbols represent 
the case $\eta=f(\alpha)$ for $\alpha=0.1$, while the open ones $\eta=1$. The solid horizontal 
line in each panel refers to the observationally inferred precession period, while 
the vertical solid line mark the location of the outer disk radius. The allowed range for the 
precession period is represented by the connecting lines, whose extremes are given by the lower 
and upper limits of the surface density of the accretion disk at the gravitational radius. Model 
predictions beyond the outer disk radius (not physically acceptable) are plotted as dashed 
lines.}
\label{RadInstability_Galactic_alpha=0.1}
\end{figure*}

\clearpage

\begin{figure*}
%%\epsscale{.80}
\plotone{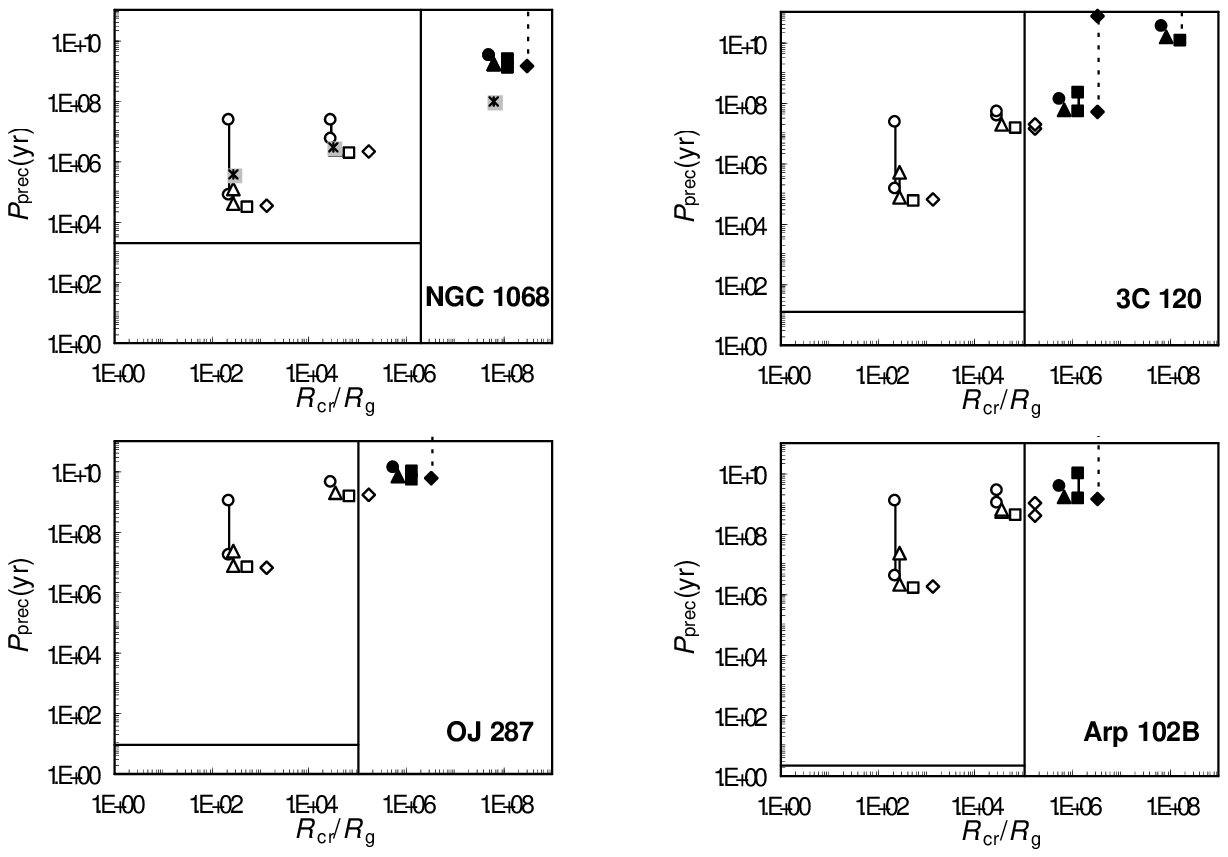}
\caption{Precession due to the radiation torques applied to our sample of AGNs. Same 
nomenclature as in Figure 2 is assumed here. In the case of $\eta=f(\alpha)$, we have 
chosen $0.001\leq\alpha\leq 0.012$ for NGC\,1068 \citep{cap06} and $\alpha=0.01$ for 
the other sources. The crossed gray squares in the panel of NGC\,1068 refer to the 
modeling of the accretion disk by \citet{hure02} based on maser data. The solid horizontal 
line refers to the observationally inferred precession period. The vertical lines 
refer to the disks' outer radii, 1.1 pc for NGC\,1068 (inferred from interferometric 
maser observations \citealt{gal04}) and $10^5 R_\mathrm{g}$ for the other sources 
(upper limit based on \citealt{codu90}).}
\label{RadInstability_Extragalactic_alpha=0.1}
\end{figure*}

\clearpage

\begin{figure*}
%%\epsscale{.80}
\plotone{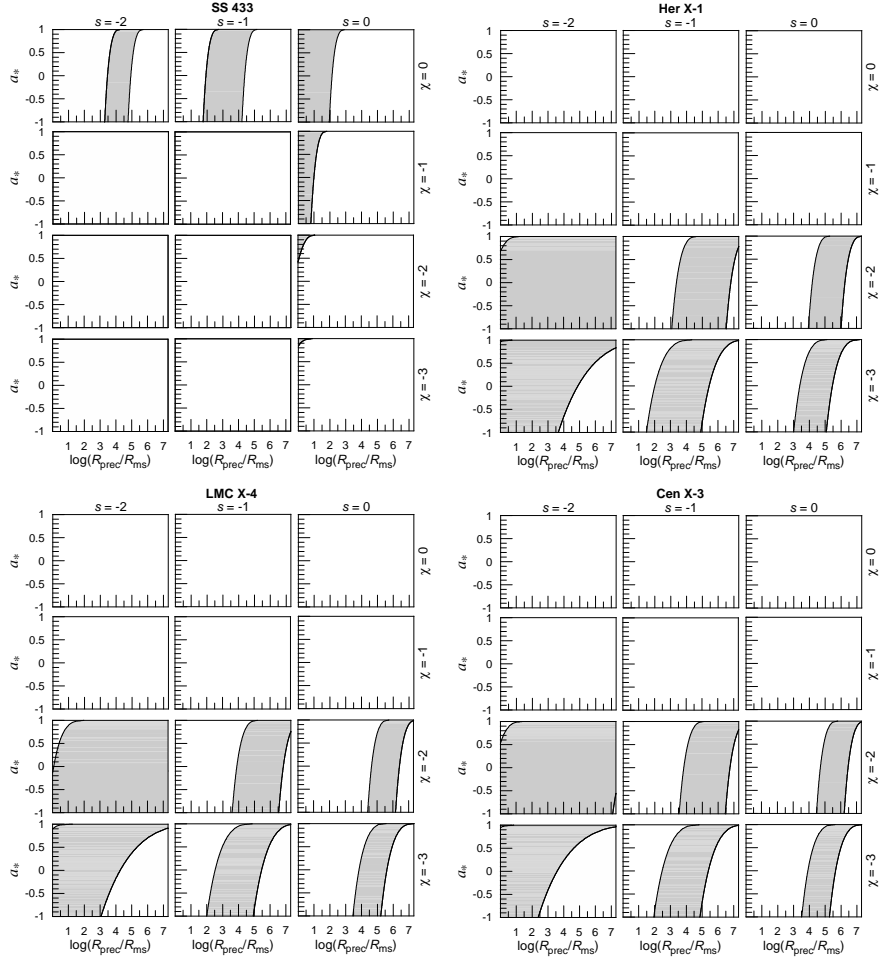}
\caption{Magnetically-driven instability. The logarithm of $\tan\varphi P_\mathrm{prec}^\mathrm{mag}B_\mathrm{0,Z}^2/\Sigma_0$ 
is calculated as a function of the dimensionless spin parameter $a_\ast$ and the outer radius of the warped/precessing part of 
the disk $R_\mathrm{prec}$ (normalized by the radius of the marginally stable orbit) for SS\,433, Her\,X-1, LMC\,X-4 and Cen\,X-3. 
$P_\mathrm{prec}^\mathrm{mag}$ is given in years, while $B_\mathrm{0,Z}$ and $\Sigma_0$ are in CGS units. The gray region in 
the panels correspond to the allowed range constrained by the observations considering $\tan\varphi=1$. Calculations were performed 
(see text) for $s=-2$, -1 and 0 (from the left to the right side) and for $\chi=0$, -1, -2 and -3 (from the top to the bottom panel).}
\label{Magnetically-driveninstability_1}
\end{figure*}

\clearpage

\begin{figure*}
%%\epsscale{.80}
\plotone{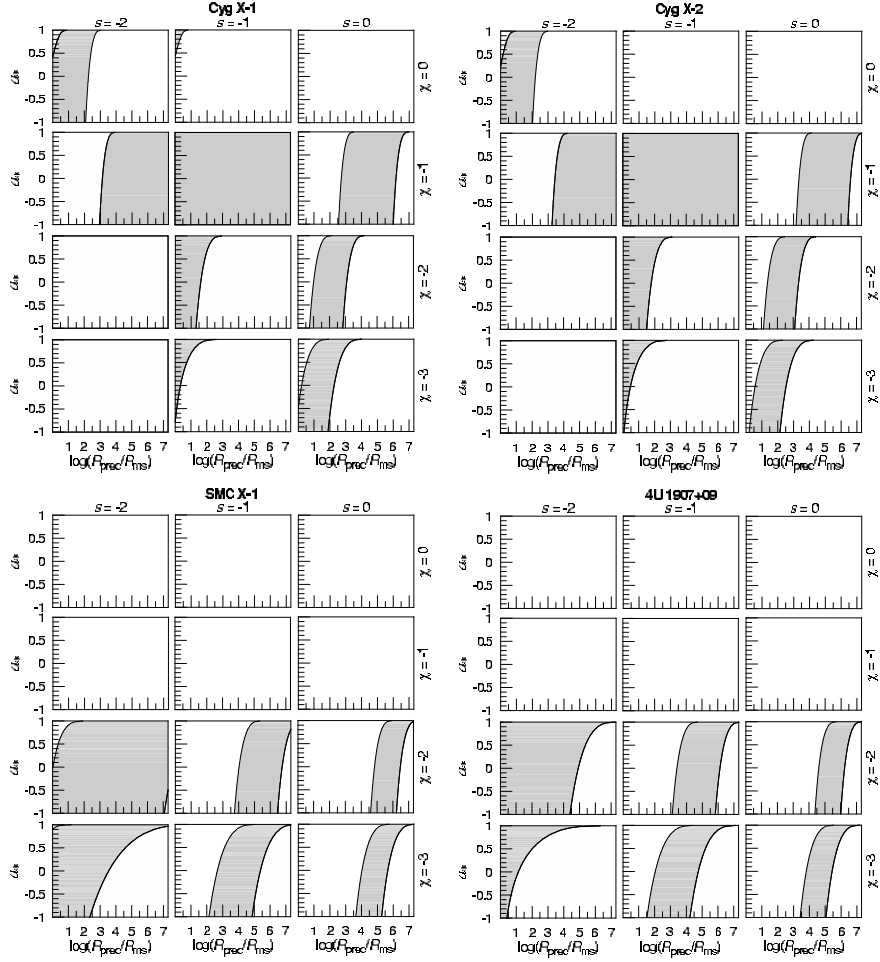}
\caption{Magnetically-driven instability for the galactic sources Cyg\,X-1, Cyg\,X-2, SMC\,X-1 and 4U\,1907+09. 
Same nomenclature as in Figure 4 is adopted here.}
\label{Magnetically-driveninstability_2}
\end{figure*}

\clearpage

\begin{figure*}
%%\epsscale{.80}
\plotone{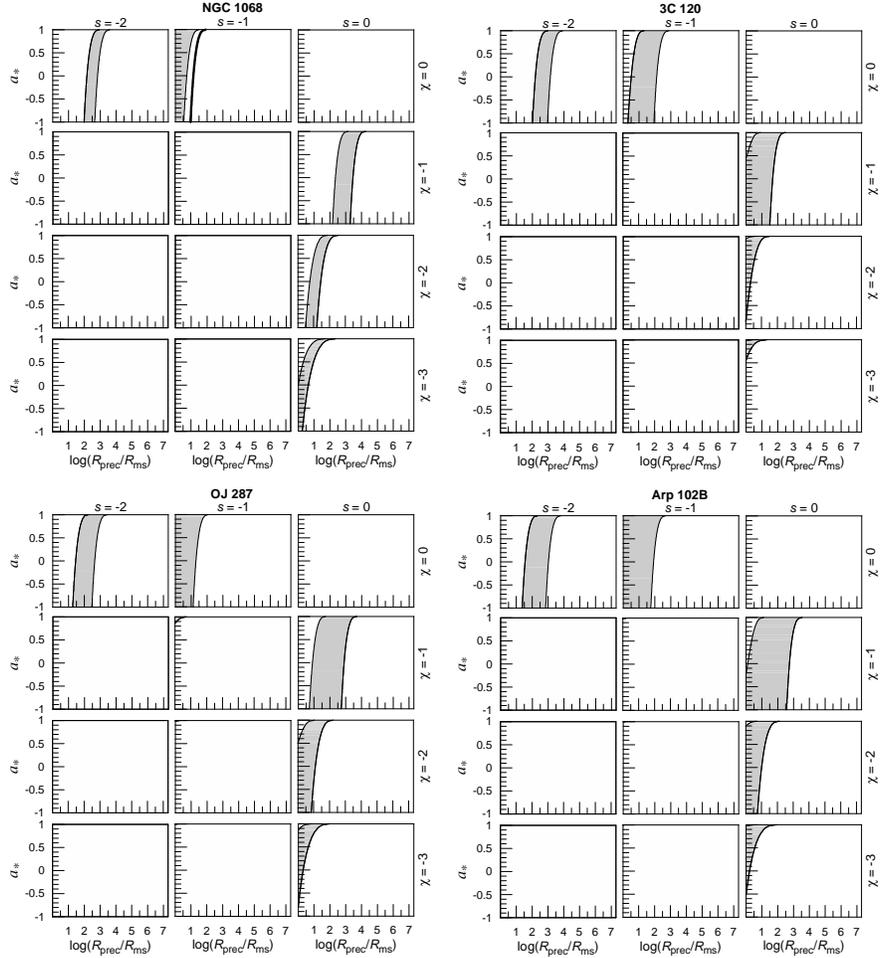}
\caption{Magnetically-driven instability for the extragalactic sources NGC\,1068, 3C\,120, OJ\,287 and Arp\,102B. 
Same nomenclature as in Figure 4 is adopted here. The thick black line represents the solution for NGC\,1068 
found from the accretion disk model parameters given by \citet{hure02} (see also \citealt{cap06}).}
\label{Magnetically-driveninstability_3}
\end{figure*}

\clearpage

\begin{figure*}
%%\epsscale{.80}
\plotone{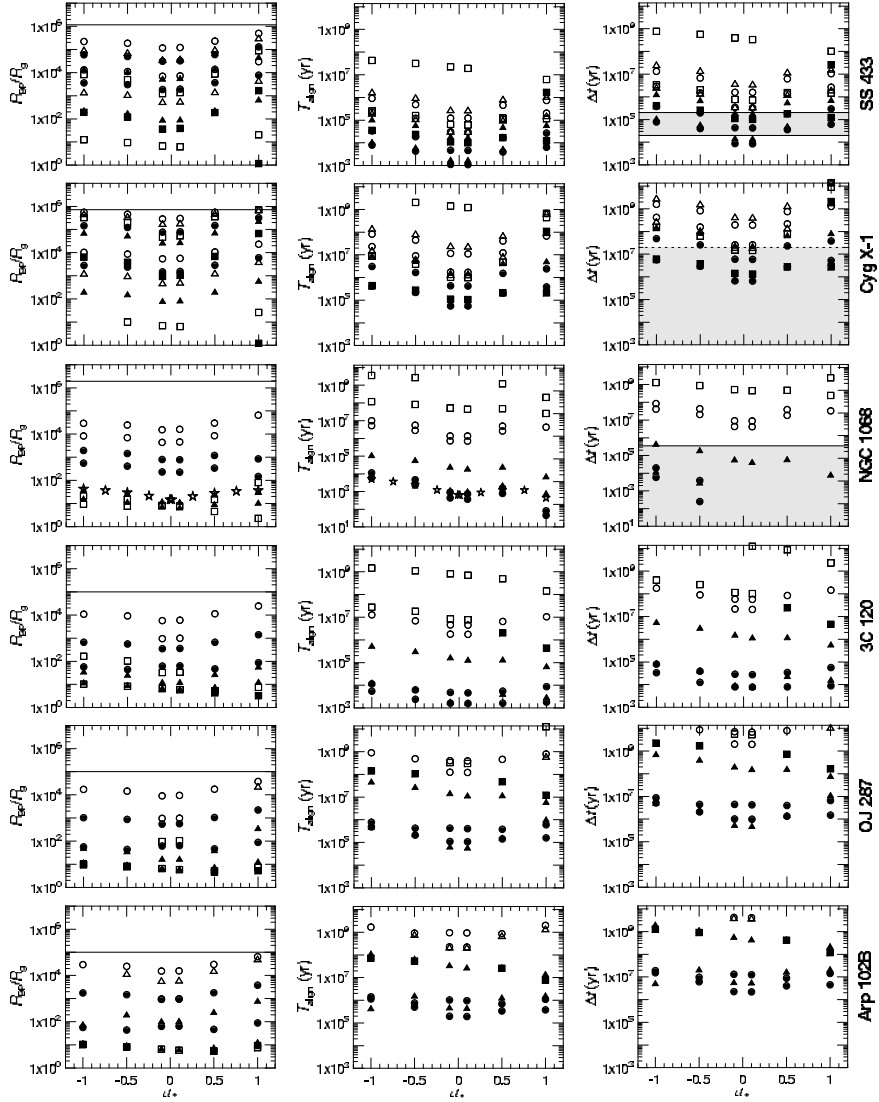}
\caption{Bardeen-Petterson effect in accreting black holes. We show the Bardeen-Petterson 
radius, the timescale for the alignment between the angular momenta 
of the black hole and of the accretion disk, and the time interval for the system to reach the observationally inferred 
precession period. Circles, triangles and squares represent respectively the solutions for $s=-2$, -1 and 0. 
Filled symbols correspond to $\eta=f(\alpha)$ ($\alpha=0.1$, for SS\,433 and Cyg\,X-1, and $\alpha=0.01$ 
for the AGN sources) while the open ones are for $\eta=1$. Stars in the panels of NGC\,1068 show the results from 
the disk model parameters given by \citet{hure02} (see also \citealt{cap06}). The horizontal lines in 
the panels on $R_\mathrm{BP}$ show the values of the disk outer radius. The two parallel lines in the $\Delta t$-plot 
for SS\,433 are the estimated age of SS\,433 (between 0.02 and 0.2 Myr; \citealt{zea80}), while 
the dashed line seen in the $\Delta t$-plot for Cyg\,X-1 is the approximated upper limit for the ages of HMXBs \citep{heu94}. 
The upper limit for age of NGC\,1068 is $3.5\times10^5$ yr \citep{wiul87}. All plots referring 
to timescales are truncated at 13.7 Gyr.}
\label{B-P Black holes}
\end{figure*}

\clearpage

\begin{figure*}
%%\epsscale{.80}
\plotone{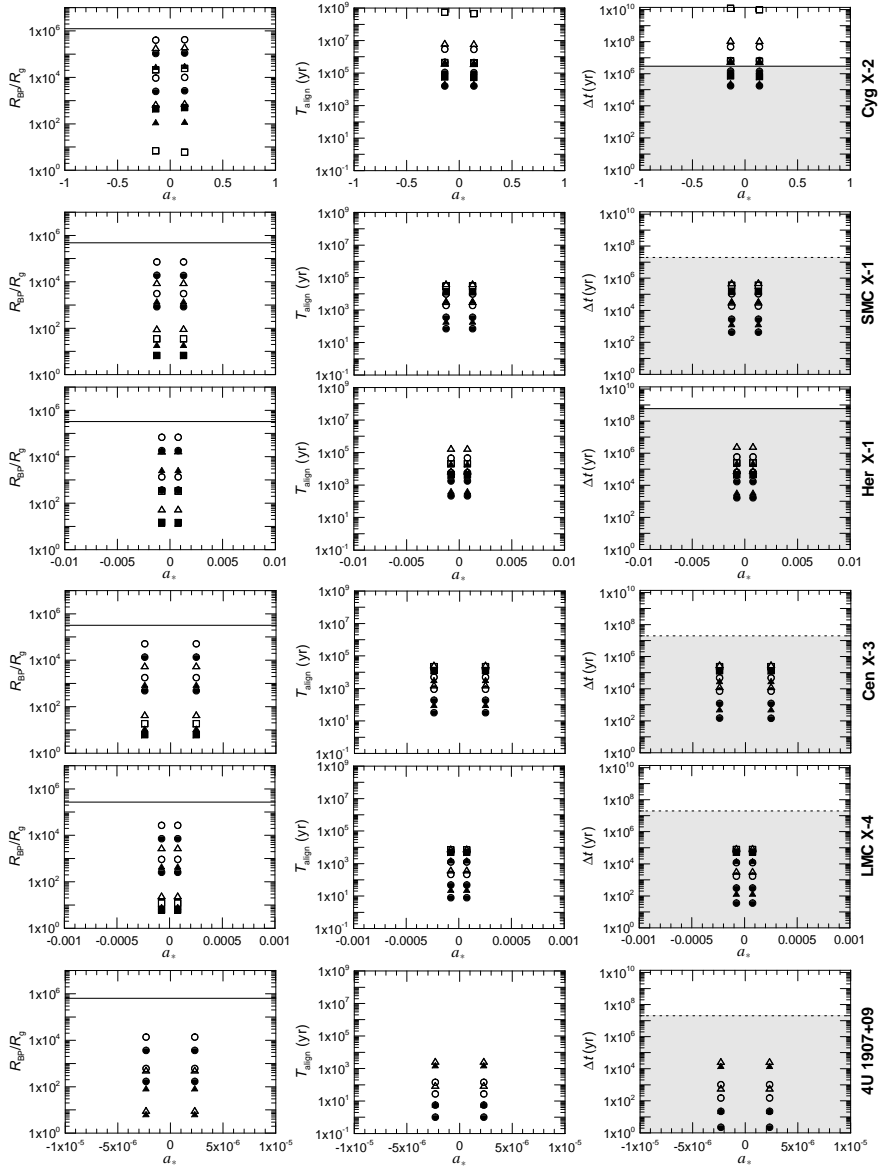}
\caption{Bardeen-Petterson effect in accreting neutron star in X-ray binaries. Same nomenclature as in Figure 7 is 
adopted here. The lower limit for the age of Her\,X-1 is $\sim 6\times10^8$ yr \citep{ver90}, while the estimated 
age of Cyg\,X-2 is roughly 3 Myr \citep{kol00}.}
\label{B-P Neutron Stars}
\end{figure*}

\end{document}